\documentclass[12pt]{article}
\usepackage{graphicx}

\newcommand\pubnumber{}
\newcommand\pubdate{\today}

\def\rwth{[1] Institute for Theoretical Particle Physics and Cosmology (TTK), \\ RWTH Aachen University, D-52056 Aachen, Germany.}
\def\lapth{[2] LAPTh, Universit\'e Savoie Mont Blanc \& CNRS, BP 110,\\ F-74941 Annecy-le-Vieux Cedex, France.}

\def\Title#1{\begin{center} {\Large #1 } \end{center}}
\def\Author#1{\begin{center}{ \sc #1} \end{center}}
\def\Address#1{\begin{center}{ \it #1} \end{center}}

\newcommand\pubblock{\rightline{\begin{tabular}{l} \pubnumber\\
         \pubdate  \end{tabular}}}
\newenvironment{Abstract}{\begin{quotation}  }{\end{quotation}}
\newenvironment{Presented}{\begin{quotation} \begin{center} 
             PRESENTED AT\end{center}\bigskip 
      \begin{center}\begin{large}}{\end{large}\end{center} \end{quotation}}

\def\beq{\begin{equation}}
\def\eeq#1{\label{#1}\end{equation}}
\def\eeqn{\end{equation}}

\def\beqa{\begin{eqnarray}}
\def\eeqa#1{\label{#1}\end{eqnarray}}
\def\eeqan{\end{eqnarray}}

\let\bar=\overbar

\def\Dslash{\not{\hbox{\kern-4pt $D$}}}
\def\dslash{\not{\hbox{\kern-2pt $\del$}}}

\def\msb{{\bar{\ssstyle M \kern -1pt S}}}

\begin{document}
\begin{titlepage}
\pubblock

\vfill
\Title{Neutrino properties from cosmology}
\vfill
\Author{Maria Archidiacono$^1$}  
\Author{Thejs Brinckmann$^1$}  
\Author{Julien Lesgourgues$^1$}  
\Author{Vivian Poulin$^{1,2}$}  
\Address{\rwth}
\Address{\lapth}
\vfill
\begin{Abstract}
The interplay between cosmology and earth based experiments is crucial in order to pin down neutrino physics.
Indeed cosmology can provide very tight, yet model dependent, constraints on some neutrino properties.
Here we focus on the neutrino mass sum, reviewing the up to date current bounds and showing the results of our forecast of the sensitivity of future experiments. Finally, we discuss the case for sterile neutrinos, explaining how non standard sterile neutrino self-interactions can reconcile the oscillation anomalies with cosmology.
\end{Abstract}
\vfill
\begin{Presented}
NuPhys2016, Prospects in Neutrino Physics\\
Barbican Centre, London, UK,  December 12--14, 2016
\end{Presented}
\vfill
\end{titlepage}
\def\thefootnote{\fnsymbol{footnote}}
\setcounter{footnote}{0}

\section{Introduction}

One of the cornerstone of particle cosmology is the possibility of constraining neutrino physics using cosmological data. The impact of neutrinos on cosmological observables has been widely studied in the literature~\cite{Bashinsky:2003tk,Hannestad:2010kz,Lesgourgues:2012uu,Lesgourgues:1519137,
Archidiacono:2013fha}. However it is important to stress the complementarity between cosmological constraints and the results of neutrino experiments. Indeed, cosmology is sensitive only to the neutrino mass sum $M_\nu=\Sigma m_\nu$ and to the number of relativistic degrees of freedom $N_\mathrm{eff}$. Therefore, the final answer to open questions such as the Dirac vs. Majorana nature, can come only from earth based neutrino experiments.

The remainder of the article is organized as follows: in Section~\ref{sec:mnu} we review the current cosmological upper bounds on the neutrino mass sum and we present a forecast of the sensitivity of future CMB experiments and galaxy surveys. In Section~\ref{sec:neff} we discuss how  the current constraints on the effective number of relativistic degrees of freedom do not leave room for sterile neutrinos in cosmology unless new physics comes into play. Finally, we draw our conclusions in Section~\ref{sec:conclusions}.

\section{Massive neutrinos}
\label{sec:mnu}

Massive neutrinos can be considered as a hot dark matter component because they are relativistic when they decouple from the thermal bath ($T \sim 1$~MeV).
Cosmology can measure the hot dark matter density, which is related to the neutrino mass though the following formula\footnote{It has to be stressed that this formula does not account for distortions in the neutrino distribution.}:
\begin{equation}
\Omega_\nu h^2=\frac{M_\nu}{93.14 \, \mathrm{eV}}.
\end{equation}

The impact of massive neutrinos on the evolution of the universe is closely related to their mass, i.e. the time of their non relativistic transition: before that neutrinos behave as radiation, while after they represent an additional matter component. This time dependent phenomenology affects the constraints that can be derived from observables located at different redshift.

\subsection{Current bounds}

If neutrinos have an individual mass smaller than $\sim 0.6$~eV, they are still relativistic at the time of photon decoupling (last scattering surface, $z \sim 1100$). Therefore the Cosmic Microwave Background (CMB) is not the best probe to constrain the neutrino mass. Indeed, the background and perturbation effects of a non zero neutrino mass on the CMB temperature anisotropy power spectrum (TT) can be mimicked by a variation of other cosmological parameters (see Section~\ref{sec:mnu} for a discussion of the degeneracy problem).  
Current bounds from Planck (TT only) indicates a neutrino mass smaller than $0.59$~eV at $95$\% c.l.~\cite{Aghanim:2016yuo}.

This bound strongly improves once CMB lensing is taken into account ($M_\nu<0.14$~eV at $95$\% c.l.). The reason is that CMB lensing probes the deflection of CMB photons as they travel from the last scattering surface towards us. Thus, CMB lensing is sensitive to the matter distribution at intermediate redshift ($0.1<z<5$), when massive neutrinos are already non relativistic.

After the non relativistic transition, neutrinos start free streaming, i.e., because of their high velocity dispersion, they cannot cluster on scales smaller than the so called free streaming scale. 
The effect on structure formation is twofold: on one hand they do not cluster themselves and on the other hand they slow down the growth of cold dark matter perturbations. The imprint of these effects can be identified in low redshift galaxy surveys, which provide both a geometric information (the Baryonic Acoustic Oscillation (BAO) scale) and a shape information (the overall matter power spectrum $P(k)$). The latter information is more efficient in constraining the neutrino mass sum, because it is sensitive to the characteristic suppression of small scale clustering due to free streaming. However, the shape information is more prone to systematics and, in particular, to the uncertainty on non linear effects. Including both the BAO and $P(k)$ information extracted from the Sloan Digital Sky Survey Data Release 7, the authors of Reference~\cite{Cuesta:2015iho} find $M_\nu<0.13$~eV at $95$\% c.l. (without CMB lensing).
Besides BAO and $P(k)$, a further source of information about the matter distribution is provided by
galaxy lensing: the distortion of the images of distant galaxies by intervening matter. Lensing data from the Canada France Hawaii Lensing Survey (CFHTLenS), together with Planck CMB temperature and lensing datasets and BAO, lead to $M_\nu<0.30$~eV at $95$\% c.l. \cite{Leistedt:2014sia, Beutler:2014yhv}.

\subsection{Future constraints}

\begin{figure}[h]
\begin{tabular}{ll}
\includegraphics[height=2in]{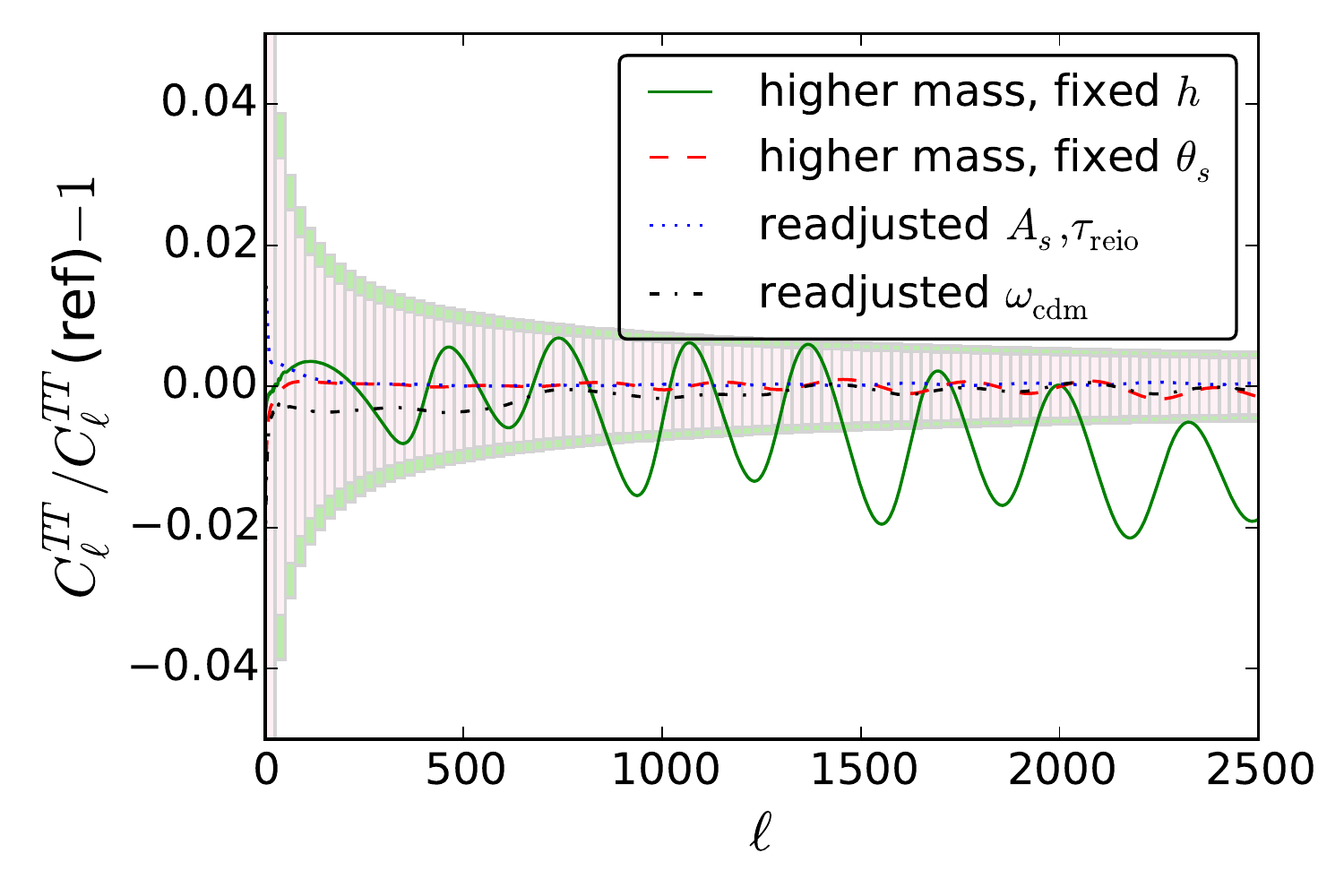}&
\includegraphics[height=2in]{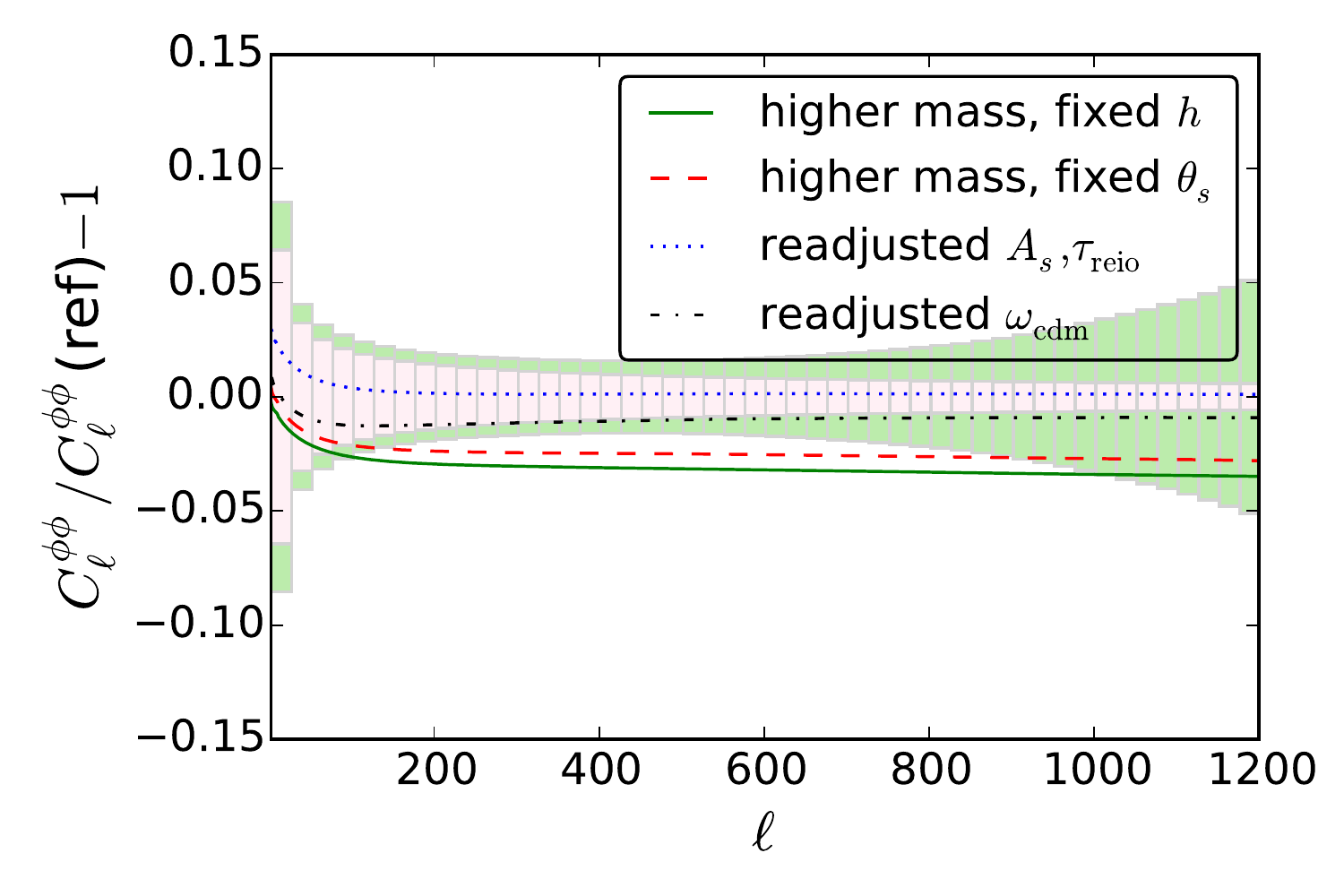}
\end{tabular}
\caption{Relative change in the CMB spectra induced by increasing the summed neutrino mass from $M_\nu=60$~meV to $M_\nu=150$~meV. The plots shows the residuals of the lensed $TT$ (left) and lensing potential (right) power spectrum, as a function of multipoles $\ell$. The light/pink and darker/green shaded rectangles refer, respectively, to the binned noise spectrum of a cosmic-variance-limited or CORE-like experiment, with linear bins of width $\Delta \ell=25$. 
The physical baryon density $\omega_b$ and the scalar spectral index $n_s$ are kept fixed. In the first case (green solid line) the value of the Hubble constant is fixed at the reference value, while in all the other cases (labeled as fixed $\theta_s$) $h$ decreases in order to keep $\theta_s$ consistent with the reference model. Moreover, in the third case (dotted blue line), we tried to compensate for the changes in the lensing spectrum by increasing $A_s$, and in the fourth case (dotted-dashed black) we aim at the same result by increasing $\omega_\mathrm{cdm}$.}
\label{fig:TT}
\end{figure}

As we already mentioned, the effect of a larger neutrino mass can be compensated by a variation of other cosmological parameters. Degeneracies among cosmological parameters limit the sensitivity of the data to the parameters. The effect is shown in Figure~\ref{fig:TT} for the sensitivity of a future CORE-like CMB experiment~\cite{Finelli:2016cyd} to the neutrino mass sum.
Increasing the neutrino mass sum from $60$~meV to $150$~meV and keeping all the other parameters constant, apparently induces a detectable effect (green solid line), both in the TT spectrum (left panel) and in the lensing spectrum (right panel). However, if, besides increasing the neutrino mass, we adjust the reduced Hubble constant $H_0=h\,100$~km/s/Mpc, so that the sound horizon angular scale at photon decoupling $\theta_s$ remains constant (red dashed line), then the effect is within the observational error (including cosmic variance) in the TT spectrum (left panel). If, at the same time, we also readjust the amplitude of the primordial power spectrum $A_s$ and the reionization optical depth $\tau_\mathrm{reio}$ (blue dotted line), or the cold dark matter density $\omega_\mathrm{cdm}$ (black dot dashed line), then the effect of the increased neutrino mass is completely swamped in the lensing spectrum as well (right panel).

It should be stressed that the model here is the minimal $\Lambda$CDM plus massive neutrinos, i.e. the parameter space is 
$\left\lbrace \omega_b, \omega_{\rm cdm}, h,n_s, A_s,\tau_{\rm reio},M_\nu  \right\rbrace$. In the case of an extended cosmological model, e.g. with a varying dark energy equation of state parameter $w$, the degeneracies would be even more severe and the constraints on $M_\nu$ would degrade.

In order to assess the sensitivity of a future CORE-like CMB experiment to the neutrino mass sum, in Reference~\cite{Archidiacono:2016lnv} we perform a Markov Chain Monte Carlo (MCMC) forecast, using the {\sc MontePython} package\footnote{\tt http://baudren.github.io/montepython.html}~\cite{Audren:2012wb}, interfaced with the Boltzmann solver {\sc class}\footnote{\tt http://class-code.net}~\cite{Lesgourgues:2011re,Blas:2011rf,Lesgourgues:2011rh}. For a fiducial neutrino mass sum of $60$~meV, close to the minimum allowed value in the normal mass ordering, the sensitivity is $\sigma(M_\nu)=42$~meV.

\begin{figure}[h]
\begin{tabular}{ll}
\includegraphics[height=2.3in]{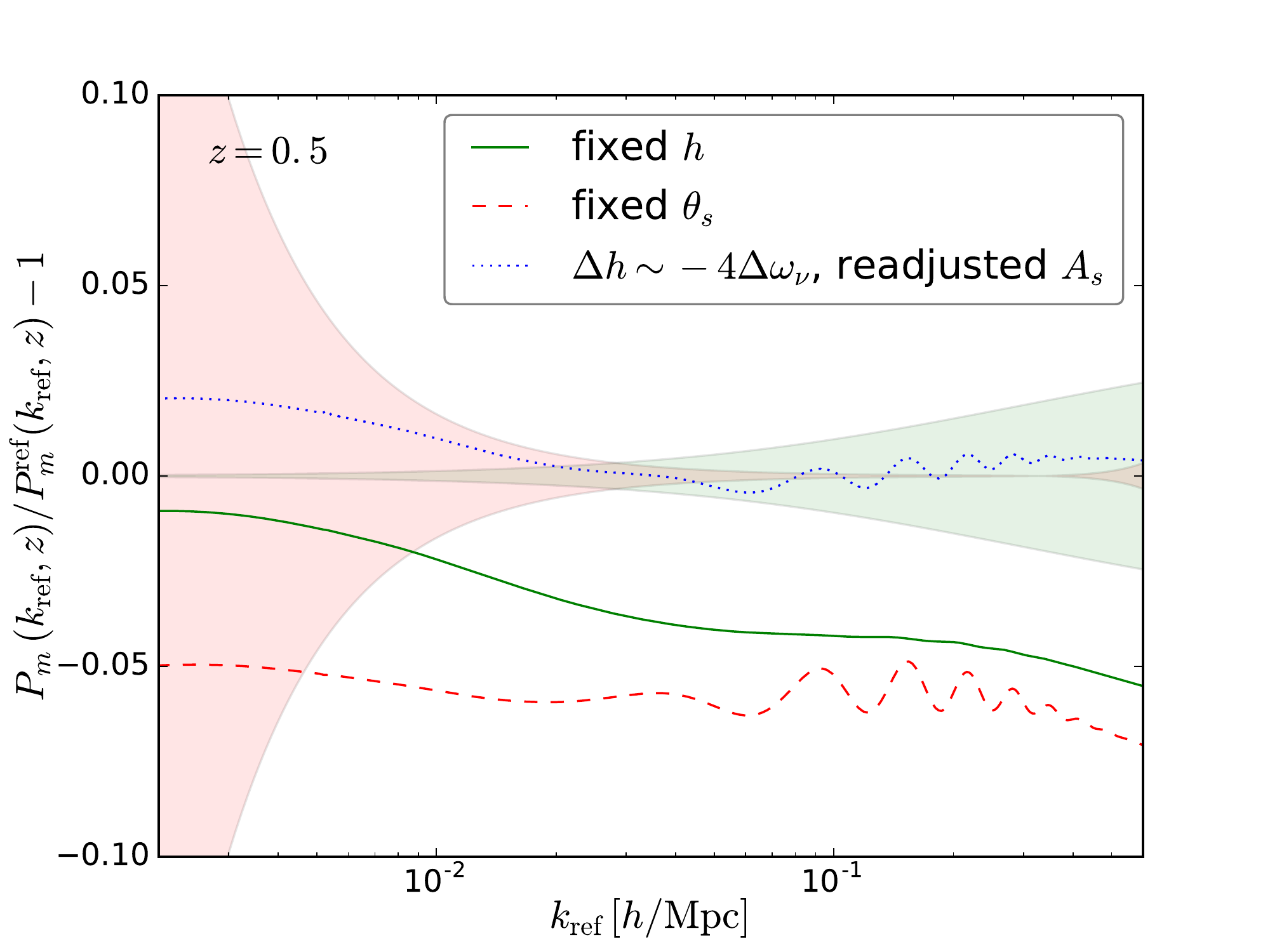}&
\includegraphics[height=2.3in]{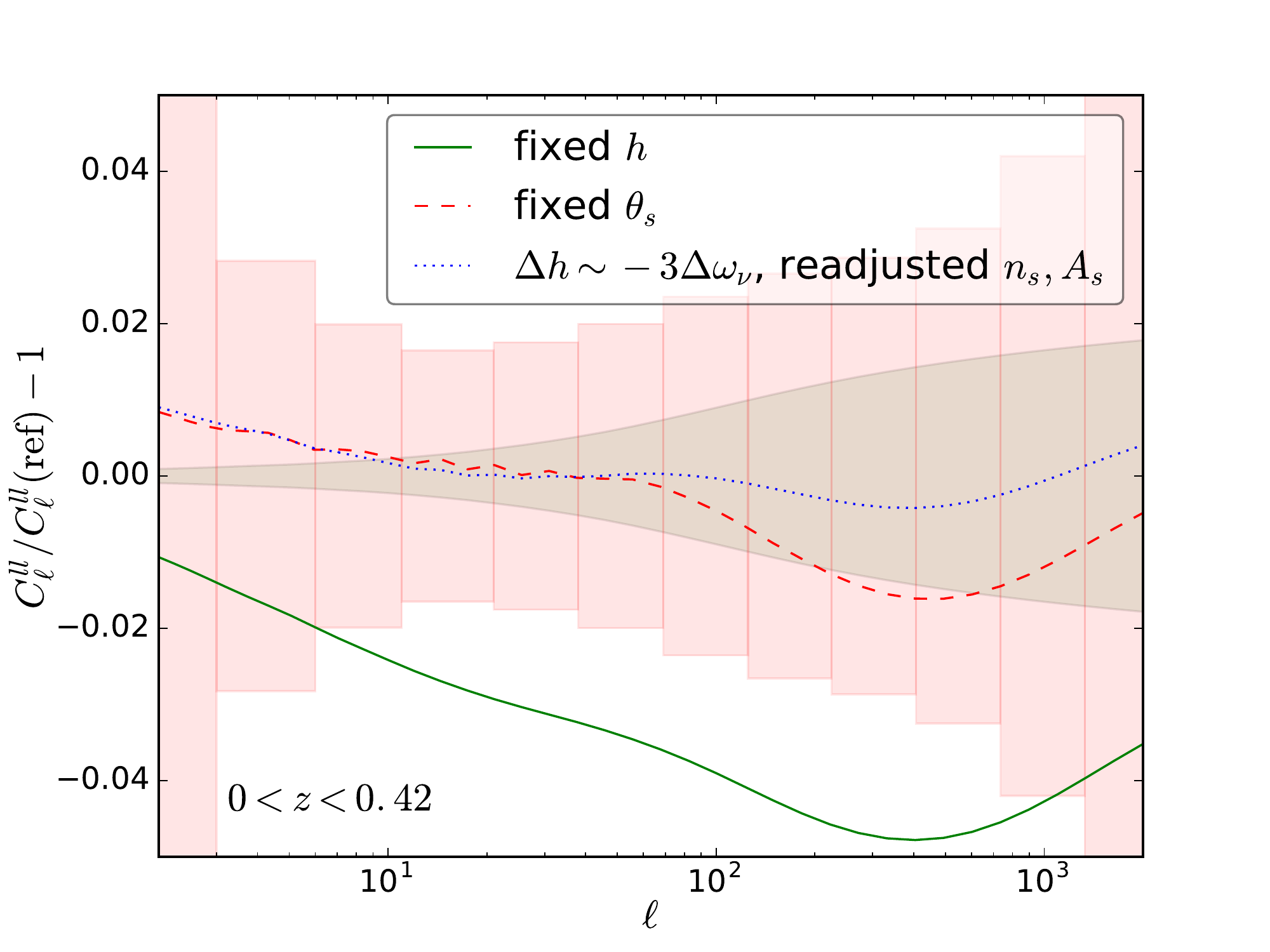}
\end{tabular}
\caption{
{\it Left panel}: Relative error on the 3D matter power spectrum $P_m(k)$ perpendicular to the line of sight at redshift $z=0.5$. Here the redshift range is $0.5<z<2$ and is divided in 16 redshift bins. 
{\it Right panel}: Relative error on the 2D galaxy lensing angular power spectrum $C_\ell^{ll}$ in the first redshift bin. Here the redshift range is $0<z<2.5$ and is divided in ten equi-populated redshift bins.
The light pink shaded area refers to the observational error, including cosmic variance.
The green shaded area represents a global uncorrelated theoretical error on non linear scales.
Green solid line and red dashed lines are the same as in figure~\ref{fig:TT}, i.e. higher $M_\nu$ with fixed $h$ (green solid line) and higher $M_\nu$ with fixed $\theta_s$ and varying $h$ (red dashed line). The blue dotted line, besides the higher $M_\nu$, also implies a smaller value of $h$ and an increase of $n_s$ and of $A_s$.}
\label{fig:Pk}
\end{figure}

In order to improve this sensitivity to the neutrino mass sum, it is crucial to disentangle the effect of different parameters by taking into account datasets from different redshift.
Therefore, including observables such as galaxy clustering and cosmic shear, extracted from a future Euclid-like galaxy survey, greatly improves the cosmological sensitivity to the neutrino mass sum. The degeneracy breaking mechanism is illustrated in Figure~\ref{fig:Pk}.
While in the TT spectrum the increase of the neutrino mass could be compensated by fixing $\theta_s$, i.e. decreasing $h$ (red dashed line in Figure~\ref{fig:TT}, left panel), the same adjustment does not work in the matter power spectrum (red dashed line in Figure~\ref{fig:Pk}, left panel). In order to compensate for the effect of massive neutrinos in galaxy clustering and shear spectra, other parameters (e.g. $A_s,n_s$) must be varied. Precisely because of this degeneracy breaking mechanism, a joint forecast of a CORE-like CMB experiment and a Euclid-like galaxy survey improves the sensitivity to the neutrino mass sum: $\sigma(M_\nu)=14$~meV.

Finally, a very promising technique to shed light on the evolution of the Universe at intermediate redshift is the detection of the redshifted 21 cm line with radio telescopes, such as the future Square Kilometer Array (SKA)~\cite{Mao:2008ug}. References~\cite{Villaescusa-Navarro:2015cca,Liu:2015txa} have shown that the independent measurement of the epoch of reionization by 21cm surveys could improve the determination of the optical depth to reionization\footnote{Notice that the constraints on $\tau_\mathrm{reio}$ from 21 cm surveys are strongly affected by astrophysical uncertainties.}, and thus of the summed neutrino mass. To assess the impact of 21cm surveys on $\sigma(M_\nu)$,
we combine the CORE plus Euclid forecast with a gaussian prior on $\tau_{\rm reio}$ ($\sigma(\tau_{\rm reio}) = 0.001$~\cite{Santos:2015gra,Liu:2015txa}). In this case the sensitivity to a fiducial neutrino mass sum of $60$~meV becomes $\sigma(M_\nu)=12$~meV.

\section{Sterile neutrinos and new physics}
\label{sec:neff}

The effective number of relativistic degrees of freedom $N_\mathrm{eff}$ accounts for any species, besides photons, that are relativistic in the early Universe. In the standard three neutrino flavor scenario $N_\mathrm{eff}^\mathrm{SM}=3.046$ \cite{Mangano:2005cc}. Planck constraints on $N_\mathrm{eff}$ ($N_\mathrm{eff}=3.15 \pm 0.23$~\cite{Ade:2015xua}) are consistent with the standard value. Light sterile neutrinos, as hinted at by oscillation anomalies, would contribute to $N_\mathrm{eff}$ with an additional $\Delta N_\mathrm{eff}=1$ in the 3+1 model ($\Delta N_\mathrm{eff}=2$ in the 3+2 model ). Therefore, eV sterile neutrinos are in stark contrast with cosmology, not only because of the constraints on $N_\mathrm{eff}$, but also because of the bounds on the hot dark matter density (see Section~\ref{sec:mnu}, $M_\nu<0.13$~eV at $95$\% c.l.). 

In order to accommodate sterile neutrinos in cosmology, we need to introduce some new physics. Non-standard sterile neutrino self-interactions mediated by a new boson can play this role. The interactions delay the sterile neutrino production until after the active neutrino collisional decoupling. Thus, once the sterile states are populated, they are not thermally distributed and their contribution to $N_\mathrm{eff}$ is smaller~\cite{Archidiacono:2014nda, Archidiacono:2015oma, Chu:2015ipa, Archidiacono:2016kkh, Cherry:2016jol, Forastieri:2017oma}.

Moreover, if the mediator is a light pseudoscalar with a mass much smaller than the neutrino mass, sterile neutrinos and the mediator recouple at late times, before sterile neutrinos go non relativistic. After the non relativistic transition, slightly before photon decoupling, sterile neutrinos start annihilating into pseudoscalars. As a consequence, the sterile neutrino mass has no visible effect on the cosmological observables. This model makes sterile neutrinos consistent with the cosmological bounds both on $N_\mathrm{eff}$ and on $M_\nu$.

\section{Conclusions}
\label{sec:conclusions}

Cosmological constraints on neutrino properties are limited to some specific parameters, namely the hot dark matter density and the effective number of relativistic degrees of freedom, which, under some assumptions about the neutrino phase space distribution, can be interpreted as the neutrino mass sum and the number of neutrino species, respectively.

Concerning the neutrino mass sum, current cosmological upper bounds are by far more stringent than laboratory limits. However, any cosmological result is model dependent. This means that in a (plausible) extended cosmological model the cosmological upper bounds on $M_\nu$ would be relaxed. Our MCMC forecast of the sensitivity of future CMB experiments and galaxy surveys show that a neutrino mass sum close to the minimum allowed value in the normal hierarchy can be detected at more than $4\,\sigma$.

Finally, concerning the effective number of relativistic degrees of freedom, we have mentioned how non-standard self-interactions can reconcile eV sterile neutrinos in cosmology.



\begin{thebibliography}{99}


\bibitem{Bashinsky:2003tk}
  S.~Bashinsky and U.~Seljak,
  Phys.\ Rev.\ D {\bf 69} (2004) 083002
  doi:10.1103/PhysRevD.69.083002
  [astro-ph/0310198].

\bibitem{Hannestad:2010kz}
  S.~Hannestad,
  Prog.\ Part.\ Nucl.\ Phys.\  {\bf 65} (2010) 185
  doi:10.1016/j.ppnp.2010.07.001
  [arXiv:1007.0658 [hep-ph]].

\bibitem{Lesgourgues:2012uu}
  J.~Lesgourgues and S.~Pastor,
  Adv.\ High Energy Phys.\  {\bf 2012} (2012) 608515
  doi:10.1155/2012/608515
  [arXiv:1212.6154 [hep-ph]].

\bibitem{Lesgourgues:1519137}
  J.~Lesgourgues, G.~Mangano, G.~Miele,
                       and S.~Pastor,
    Cambridge Univ. Press" (2013)

\bibitem{Archidiacono:2013fha}
  M.~Archidiacono, E.~Giusarma, S.~Hannestad and O.~Mena,
  Adv.\ High Energy Phys.\  {\bf 2013} (2013) 191047
  doi:10.1155/2013/191047
  [arXiv:1307.0637 [astro-ph.CO]].
  
 
\bibitem{Aghanim:2016yuo}
  N.~Aghanim {\it et al.} [Planck Collaboration],
  Astron.\ Astrophys.\  {\bf 596} (2016) A107
  doi:10.1051/0004-6361/201628890
  [arXiv:1605.02985 [astro-ph.CO]].
  
\bibitem{Cuesta:2015iho}
  A.~J.~Cuesta, V.~Niro and L.~Verde,
  Phys.\ Dark Univ.\  {\bf 13} (2016) 77
  doi:10.1016/j.dark.2016.04.005
  [arXiv:1511.05983 [astro-ph.CO]].
  
\bibitem{Leistedt:2014sia}
  B.~Leistedt, H.~V.~Peiris and L.~Verde,
  Phys.\ Rev.\ Lett.\  {\bf 113} (2014) 041301
  doi:10.1103/PhysRevLett.113.041301
  [arXiv:1404.5950 [astro-ph.CO]].
  
\bibitem{Beutler:2014yhv}
  F.~Beutler {\it et al.} [BOSS Collaboration],
  Mon.\ Not.\ Roy.\ Astron.\ Soc.\  {\bf 444} (2014) no.4,  3501
  doi:10.1093/mnras/stu1702
  [arXiv:1403.4599 [astro-ph.CO]].
  
\bibitem{Archidiacono:2016lnv}
  M.~Archidiacono, T.~Brinckmann, J.~Lesgourgues and V.~Poulin,
  JCAP {\bf 1702} (2017) no.02,  052
  doi:10.1088/1475-7516/2017/02/052
  [arXiv:1610.09852 [astro-ph.CO]].
  
\bibitem{Finelli:2016cyd}
  F.~Finelli {\it et al.} [CORE Collaboration],
  arXiv:1612.08270 [astro-ph.CO].
  
\bibitem{Audren:2012wb}
  B.~Audren, J.~Lesgourgues, K.~Benabed and S.~Prunet,
  JCAP {\bf 1302} (2013) 001
  doi:10.1088/1475-7516/2013/02/001
  [arXiv:1210.7183 [astro-ph.CO]].
  
\bibitem{Lesgourgues:2011re}
  J.~Lesgourgues,
  arXiv:1104.2932 [astro-ph.IM].
  
\bibitem{Blas:2011rf}
  D.~Blas, J.~Lesgourgues and T.~Tram,
  JCAP {\bf 1107} (2011) 034
  doi:10.1088/1475-7516/2011/07/034
  [arXiv:1104.2933 [astro-ph.CO]].

\bibitem{Lesgourgues:2011rh}
  J.~Lesgourgues and T.~Tram,
  JCAP {\bf 1109} (2011) 032
  doi:10.1088/1475-7516/2011/09/032
  [arXiv:1104.2935 [astro-ph.CO]].

\bibitem{Mao:2008ug}
  Y.~Mao, M.~Tegmark, M.~McQuinn, M.~Zaldarriaga and O.~Zahn,
  Phys.\ Rev.\ D {\bf 78} (2008) 023529
  doi:10.1103/PhysRevD.78.023529
  [arXiv:0802.1710 [astro-ph]].

\bibitem{Villaescusa-Navarro:2015cca}
  F.~Villaescusa-Navarro, P.~Bull and M.~Viel,
  Astrophys.\ J.\  {\bf 814} (2015) no.2,  146
  doi:10.1088/0004-637X/814/2/146
  [arXiv:1507.05102 [astro-ph.CO]].

\bibitem{Liu:2015txa}
  A.~Liu, J.~R.~Pritchard, R.~Allison, A.~R.~Parsons, U.~Seljak and B.~D.~Sherwin,
  Phys.\ Rev.\ D {\bf 93} (2016) no.4,  043013
  doi:10.1103/PhysRevD.93.043013
  [arXiv:1509.08463 [astro-ph.CO]].

\bibitem{Santos:2015gra}
  M.~G.~Santos {\it et al.},
  PoS AASKA14 (2015) 019
  [arXiv:1501.03989 [astro-ph.CO]].

\bibitem{Mangano:2005cc}
  G.~Mangano, G.~Miele, S.~Pastor, T.~Pinto, O.~Pisanti and P.~D.~Serpico,
  Nucl.\ Phys.\ B {\bf 729} (2005) 221
  doi:10.1016/j.nuclphysb.2005.09.041
  [hep-ph/0506164].
  
\bibitem{Ade:2015xua}
  P.~A.~R.~Ade {\it et al.} [Planck Collaboration],
  Astron.\ Astrophys.\  {\bf 594} (2016) A13
  doi:10.1051/0004-6361/201525830
  [arXiv:1502.01589 [astro-ph.CO]].
  
\bibitem{Archidiacono:2014nda}
  M.~Archidiacono, S.~Hannestad, R.~S.~Hansen and T.~Tram,
  Phys.\ Rev.\ D {\bf 91} (2015) no.6,  065021
  doi:10.1103/PhysRevD.91.065021
  [arXiv:1404.5915 [astro-ph.CO]].
  
\bibitem{Archidiacono:2015oma}
  M.~Archidiacono, S.~Hannestad, R.~S.~Hansen and T.~Tram,
  Phys.\ Rev.\ D {\bf 93} (2016) no.4,  045004
  doi:10.1103/PhysRevD.93.045004
  [arXiv:1508.02504 [astro-ph.CO]].

\bibitem{Chu:2015ipa}
  X.~Chu, B.~Dasgupta and J.~Kopp,
  JCAP {\bf 1510} (2015) no.10,  011
  doi:10.1088/1475-7516/2015/10/011
  [arXiv:1505.02795 [hep-ph]].

\bibitem{Archidiacono:2016kkh}
  M.~Archidiacono, S.~Gariazzo, C.~Giunti, S.~Hannestad, R.~Hansen, M.~Laveder and T.~Tram,
  JCAP {\bf 1608} (2016) no.08,  067
  doi:10.1088/1475-7516/2016/08/067
  [arXiv:1606.07673 [astro-ph.CO]].


\bibitem{Cherry:2016jol}
  J.~F.~Cherry, A.~Friedland and I.~M.~Shoemaker,
  arXiv:1605.06506 [hep-ph].
  
\bibitem{Forastieri:2017oma}
  F.~Forastieri, M.~Lattanzi, G.~Mangano, A.~Mirizzi, P.~Natoli and N.~Saviano,
  arXiv:1704.00626 [astro-ph.CO].


\end{thebibliography}
\end{document}